\documentclass[a4paper,11pt]{article} 

\usepackage{amsmath,amsfonts,amssymb} 
\usepackage{graphics,graphicx}				  

\usepackage[colorlinks=false]{hyperref}
\usepackage{latexsym}
\usepackage{flexisym,mathstyle}		  
\usepackage{breqn}    				  
\usepackage{epstopdf}   			  
\usepackage[T1]{fontenc}
\usepackage{subcaption}
\usepackage{authblk}

\title{Implications of the complex Singlet field for Noncommutative Geometry model}
\author[1]{Hosein Karimi \thanks{hmk35@mail.aub.edu}}
\affil[1]{Physics Department, American University of Beirut, Lebanon}
\date{}

\begin{document}
	
	\maketitle

\begin{abstract}
		We consider a complex singlet scalar in the spectral action approach to the standard model. It is shown that there is a range of initial values at the unification scale which is able to produce Higgs and top quark masses at low energies. The stability of the vacuum and the deviation of gauge couplings from experimental values are discussed and compared at the two-loop level with a real scalar singlet and the pure standard model.
\end{abstract}
	
	\tableofcontents
	
\section{Introduction}
	The noncommutative geometry approach to the structure of space-time has been able to produce the standard model coupled with gravity, almost uniquely, by using very weak constraints \cite{Chams-1997}. In this model, space-time is taken to consist of a continuous $4D$ Riemannian manifold tensored with a finite noncommutative space. One of the defining ingredients of this hyperspace is an operator which coincides with the Dirac operator in the commutative $4D$ part of the space and can be considered as the generalized noncommutative version of it. This operator has all the useful geometrical information of the space, and just like the Dirac operator in the standard model, its structure reveals the fermionic content of the model. Moreover, in the noncommutative geometry, other information like gauge field interactions and the scalar sector are embedded in the spectrum of this operator. In the work of Chamseddine and Connes in \cite{Chams-1997} and the papers that followed, it was shown that the simplest possible noncommutative structure has the correct fermionic content and also leads to the gauge symmetry of the standard model. 
	
	The Lagrangian of this model comes from the most general form of the Dirac operator consistent with axioms of noncommutative geometry plus an additional constrain called the first order condition. This Lagrangian possesses three important features distinguishing it from the minimal standard model. First, the couplings of the model are not totally arbitrary and there are relations between them at the unification scale. These relations are consistent with grand unified theories such as $SU(5)$ unified theory. Second, in addition to the Higgs, there is a singlet scalar field present in the spectral action.  It is shown that this field can help the situation with the low Higgs mass which is not otherwise consistent with the unification of spectral action in high energies \cite{Chams-2012}.  We will see in this letter that the results improve if the extra singlet scalar field is taken to be complex. It is also seen that such an extra scalar field can be responsible for dark matter particle \cite{Burgess:2000yq,Costa:2014qga}. Finally, right-handed neutrino appears into the picture automatically as well as its Yukawa interaction. These terms are needed to give a small mass to the left-handed neutrino by see-saw mechanism and usually are added to the standard model by hand.
	
	In \cite{Chams-2012}, the singlet scalar field was assumed to be real. Then using 1-loop renormalization group equations, it was shown that the model with the singlet can accommodate a Higgs field with the mass of order $125 GeV$. In fact, the reality condition on the singlet field is not necessary and we assume the singlet to be a complex field in this work. Our consideration shows the model in its most general form is consistent with the current experimental values of the Higgs and top quark masses. Furthermore, we use 2-loop renormalization group equations to compare the following cases: when the added singlet is a complex field, when it is real, and the pure standard model with neutrino mixing. We show that while running RG equations from unification scale toward current experimental energies, the model with added complex singlet behaves slightly better than the other two cases. Yet, like the standard model itself, one can only attain the experimentally observed gauge couplings at low energies within some percent of accuracy. This agrees with the separations of the standard model gauge couplings at the unification scale when we start from experimental values and run them upward. Subsequently we also discuss the effects of three-loop corrections.
	
	Since the discovery of the Higgs particle in 2012, researchers started to study the instability problem of the standard model effective potential more seriously (For example \cite{Degrassi:2012ry}). Although this instability cannot make the standard model unreliable, even at high energies, because of the long lifetime of the tunneling process, it still could have dramatic consequences during the inflation period \cite{Elias-2011,Buttazzo:2013uya}. It is interesting to check the effect of any modification of the model on this situation. Therefore the vacuum stability of the models coming from noncommutative geometry will be addressed and compared with the pure standard model. 
	
	We will show in this letter that even though a few extra terms are added to the RG equations due to the complex singlet field, yet their effect on the negativity of the Higgs self-coupling at high energies can be substantial. The reason we cannot predict what exactly happens for the coupling is that the experimentally unknown right-handed neutrino Yukawa coupling contributes in the RG equations as well. This coupling also plays a role in determining the Higgs and top quark masses at low energies. What we can do is to follow its effect by following RG equations down and looking at the particle masses. The proper value of right-handed neutrino Yukawa coupling - turns out to be between 0.411 and 0.455 at unification scale as we will see in section \ref{sec-3}. The resulting value for this coupling at Z-boson mass region is also between 0.517 and 0.530, while Yukawa coupling of the top quark is about 0.995. Besides, the values of scalar sector couplings are derivable in this scale from RG equations. We argue that in this acceptable range of the couplings, although vacuum instability is not cured, but the situation is improved by the presence of the complex scalar field. We use two-loop equations and near to the leading order three-loop equations to assess the loop correction effects in presence of a complex or real singlet field.
	
	We stress that the above results are not merely derivable from the standard model plus a complex singlet. The reason is that in our considerations, we use the initial conditions predicted by the spectral action approach \cite{Chams-2007}. Moreover here a neutrino coupling is present in RG equations and contributes to the values of particle masses. The form of potential is also restricted and is different from extended standard model cases with complex singlet described in the literature. In our case, the results for stability are slightly better (e.g. compare with \cite{Costa:2014qga,Chen:2012faa,Gonderinger:2012rd,EliasMiro:2012ay}).
	
\section{The model}\label{sec-model}
	
	After years of investigations by mathematicians to expand the geometrical notions to the spaces with less constraints than metric spaces, which led to many developments in various areas of mathematics, finally Alain Connes was able in 1980 to find applicable set of axioms and definitions to generalize geometrical concepts to a much broader range of spaces \cite{Connes-1980}. He also used the new geometry to define a noncommutative torus and studied its geometrical properties. Later in 1996, Ali Chamseddine and Alain Connes found an application of this new geometry in physics \cite{Chams-1997}. They assumed the space-time is a direct product of $4D$ Riemannian manifold with a noncommutative space. They also introduced the spectral action which is based on the spectrum of the Dirac operator and were able to show that the standard model arises naturally and almost uniquely from these assumptions. 
	
	Geometrical structures in noncommutative geometry are defined based on three concepts; a Hilbert space, an algebra of a given set of operators with its faithful representation on the Hilbert space, and a special operator called \textit{Dirac Operator}. These are shown to be enough to define a rich geometry and can yield features of Riemannian manifolds in expected limits\footnote{For precise definitions refer to \cite{Connes:1994yd,Chams-2010}.}.
	We have therefore a so called spectral triple which is shown by
	\[(\mathcal{A},\mathcal{H},D).\]
	As an example, for a $4D$ spinorial space-time one can consider Dirac operator to be the familiar $4$ by $4$ matrix $D=i\gamma^{\mu}\partial_{\mu}$. The Hilbert space is then the space of $1$ by $4$ spinors. In this case $\mathcal{A}$ is the algebra of $4$ by $4$ complex matrices which is a noncommutative algebra. Now one way to see the geometrical invariants such as curvature is to look at the spectrum of the Dirac operator. One can for example use heat kernel method to asymptotically expand the trace of Dirac operator \cite{Vassilevich:2003xt,Gilkey:1995}. This expansion is controlled by a scale called $\Lambda$. Doing so, the first term of the expansion turns out to be the cosmological constant and the second term gives the total curvature of space-time. Higher orders are higher powers of the geometrical invariants such as curvature and Ricci tensor.
	
	Unlike Kaluza-Klein type theories which enlarge geometry by assuming extra dimensions, here the added structure is a finite noncommutative space which possesses no space-time dimensions. In early models, finding noncommutative structures leading to the standard model was the matter of trial and error. Eventually, in \cite{Chams-1997}, the authors discovered that a noncommutative space with the algebra
\begin{equation}\label{eq-as}
	\mathcal{A_{F}}=\mathbb{C} \oplus \mathbb{H} \oplus M_{3}\left(\mathbb{C}\right)
\end{equation}
	is able to produce the standard model when it is tensored with the $4D$ space-time.	$M_{3}(\mathbb{C})$ is the algebra of 3 by 3 matrices on complex numbers, $\mathbb{H}$ is the algebra of quaternions which are represented using 2 by 2 matrices, and $\mathbb{C}$ is the algebra of complex numbers. Later on, the same authors showed that the classification of finite spaces consistent with the noncommutative geometry requirements leads almost uniquely to the same algebra \cite{Chams-200709}. They also observed that by letting the Dirac operator to have nonlinear fluctuations, the consistent algebra is	
\begin{equation}\label{eq-a}
	\mathcal{A_{F}}=\mathbb{H} \oplus \mathbb{H} \oplus M_{4}\left(\mathbb{C}\right),
\end{equation}
	which leads to the Pati-Salam unified model \cite{Chams-2013}. As an interesting breakthrough in 2014 it was discovered in \cite{Chams-2014} that this algebra is dictated by a generalized version of Heisenberg commutation relations. 
	In this letter we consider the model based on the algebra (\ref{eq-as}), which is a special case of (\ref{eq-a}) that happens when the perturbations of Dirac operator is required to be linear. This is called \textit{first order condition} and we assume its validity in the current work.
	
	Members of $\mathcal{A_{F}}$ are $2\times2\times4=16$ by $16$ matrices and members of the Hilbert space consist of $16$ spinors, which means they possess $64$ elements. Algebra of the whole space can be written as direct product of  $\mathcal{A_{F}}$ with the algebra of functions on the $4D$ spin manifold. The latter is the commutative algebra of smooth functions on the spin manifold
\begin{equation}\label{eq-aa}
	\mathcal{A}=C^{\infty}\left(M\right) \otimes \left(\mathbb{C} \oplus \mathbb{H} \oplus M_{3}\left(\mathbb{C}\right)\right).
\end{equation}
	
	We have then $16$ spinors and it turns out later that they have exactly the same interactions as fermions in one generation including four right and left handed leptons and $12$ colored right and left handed quarks. Next, one can introduce the chirality operator called $\gamma$ to enrich the algebra by grading mechanism and add antiparticles to the Hilbert space. Therefore members of the Hilbert space are now 1 by 128 matrices. Next, we can triple this space by hand to take into account the three generations of fermions. Dirac operator of the whole space is then a 384 by 384 matrix which acts on the Hilbert space and is defined as the tensorial sum of the operators on different parts:
\begin{equation}\label{eq-d}
	D=D_M \otimes 1_{{\Huge 96\times96}} + \gamma_{5} \otimes D_{F}.
\end{equation}
	The particle content of the model is therefore coming from the above settings of the noncommutative geometry. Then Dirac action provides dynamic to this fermionic part of the model\footnote{To be able to introduce an inner product and define this part of action consistently, another operator called reality operator is needed. For exact definitions refer to \cite{Chams-200709}}.
	The vector and scalar parts of the model are described by the spectral action which is the trace of Dirac operator and depends only on the sum of its eigenvalues. The action is:
\begin{equation}\label{eq-ac}
	S=\mathrm{Tr}\left(f(D/\Lambda)\right) + \langle \psi,D\psi \rangle.
\end{equation}	
	Lambda is an energy cutoff needed to make dimensionless term out of D. Function f is a source to generate physical constants such as $G_{N}$ and is required to be positive and even. 
	
	To start, first we need to make the fermionic part covariant under inner automorphisms of the Hilbert space by adding inner fluctuations of the Dirac operator under such automorphisms. The fluctuations associated to the noncommutative space are responsible for the existence of gauge fields and the Higgs. Inner fluctuations associated with the automorphisms of the continuous $4D$ manifold form Riemannian aspects of the curved $4D$ space-time. The Dirac action then contains all the fermionic interactions, just like the standard model when all the vector fields are added to Dirac operator in form of connections. On top of that, here we get the Yukawa terms and the Higgs as parts of the spectrum of Dirac operator.
	
	Next, one can use heat kernel asymptotic expansion to compute the trace. Existence of $\Lambda$ in the action is crucial so one can rely on the expansion\footnote{For the special case of Robertson-Walker metric it is shown that the expansion is valid up to energies close to the Planck order \cite{Chams-2011}}. The trace is then reduced to a series with coefficients known as \textit{Seeley deWitt coefficients} \cite{Gilkey-1975}: 
\begin{equation}\label{eq-seedew}
	\mathrm{Tr}\left(f\left(D/\Lambda\right)\right)=\textrm{Tr}\left(F\left(\left(D/\Lambda\right)^{2}\right)\right)=\sum_{n=0}^{\infty}\Lambda^{4-n}F_{4-n}a_{n}.
\end{equation}
	The function $f$ is supposed to be positive. The odd terms in the expansion vanish for manifolds without boundaries. It is equivalent to saying the square of the Dirac operator has important geometrical information in its spectrum and use a function $F$ such that $F(\alpha^2)=f(\alpha)$. The coefficients $a_{n}$ depend only on the geometrical invariants such as curvature and therefore reveal the geometrical information embedded in the Dirac operator up to the order defined by powers of $\Lambda$. Taylor coefficients $F_{4-n}$ are the spectral function derivatives at zero for $4-n<0$ and momenta of spectral function for $4-n>0$,
\begin{equation}\label{eq-moment}
	F_0=F(0), \qquad F_2=\int_{0}^{\infty}F(u)du, \qquad F_4=\int_0^{\infty}F(u)udu.
\end{equation}	
	These coefficients along with Yukawa couplings make the physical constants. For example the first one, $F_4$, is the source of Hubble constant and the third one, $F_0$, appears in the Higgs kinetic term. Normalization of this term causes $F_0$ to show up in the mass term of fermions as well as all the coupling constants which is the root of unification in this model \cite{Chams-2010}. Therefore we trust the model on high energies where the approximation of expansion \ref{eq-seedew} is expected to work well. The unification of the couplings will be then what is expected from GUT theories. Writing the renormalization group equations and running them down to experimental energies is also feasible.

	The sum in (\ref{eq-seedew}) is over even numbers, therefore the forth term is suppressed by $\Lambda^2$. We expect $\Lambda$ to be right below plank energy which is much higher that any mass in the model. Therefore it is logical to assume higher terms are irrelevant for our purposes. In addition, $F$ is expected to be a cutoff function which can control expansion of the trace.
	
	The Dirac operator for the noncommutative space defined by algebra in (\ref{eq-aa}) is [\cite{Chams-2010}]:
	{\small	
\begin{align}
		&D_{AB}=\gamma^\mu \otimes \\ \nonumber
		&\left(		
		\begin{tabular}{ c c c c c c c c c c c c c c c c }
		$D_\mu$&&&&&&&&&&&&&\\
		&\multicolumn{7}{c}{$D_\mu+ig_1B_\mu$}&&&&&&0&&\\ \\
		&&&\multicolumn{7}{c}{$(D_\mu+\frac{i}{2}g_1B_\mu)I_{2\times 2}-\frac{i}{2}g_2W_\mu^i \sigma_i$}&&&&&&\\ \\
		&&&&&\multicolumn{7}{c}{$(D_\mu-\frac{2i}{3}g_1B_\mu)I_{3\times 3}-\frac{i}{2}g_3V_\mu^a \lambda_a$}&&&&\\ \\
		&&&0&&&&\multicolumn{7}{c}{$(D_\mu+\frac{i}{3}g_1B_\mu)I_{3\times 3}-\frac{i}{2}g_3V_\mu^a \lambda_a$}&&\\ \\
		&&&&&&&&&\multicolumn{7}{c}{$(D_\mu-\frac{i}{6}g_1B_\mu)I_{6\times 6}-\frac{i}{2}g_3V_\mu^a \lambda_a I_{2 \times 2}-\frac{i}{2}g_2W_\mu^i \sigma_i I_{3 \times 3}$}		 
		\end{tabular}
		\right)
		\otimes 1_3 \\ \nonumber \\ \nonumber
		&+\gamma^5 \otimes \\ \nonumber
		&\left(
		\begin{tabular}{ c c c c c c c c c c c c c c}
		$0_3$&&0&\multicolumn{3}{c}{$(\epsilon^{ab}H_b \otimes k^{\ast \nu})_{6 \times 3}$}&0&&&0&&0&\\ \\
		0&&$0_3$&\multicolumn{3}{c}{$(\bar{H}^{a} \otimes k^{\ast e})_{6 \times 3}$}&0&&&0&&0&\\ \\
		$(\epsilon_{ab}\bar{H}^b \otimes k^{\nu})_{3 \times 6}$&\multicolumn{3}{c}{$(H_{a} \otimes k^{e})_{3 \times 6}$}&$0_6$&&0&&&0&&0&\\ \\
		0&&0&&0&&$0_9$&&&0&\multicolumn{3}{c}{$(\epsilon^{ab}H_b \delta_{i}^{j} \otimes k^{\ast u})_{18 \times 9}$}\\ \\
		0&&0&&0&&0&&&$0_9$&\multicolumn{3}{c}{$(\bar{H}^{a} \delta_i^j \otimes k^{\ast d})_{18 \times 9}$}\\ \\
		0&&0&&0&\multicolumn{3}{c}{$(\epsilon_{ab}\bar{H}^b \delta_{j}^{i} \otimes k^{u})_{9 \times 18}$}&\multicolumn{3}{c}{$(H_{a} \delta_j^i \otimes k^{d})_{9 \times 18}$}&$0_{18}$&
		\end{tabular}
		\right)
\end{align}	
}
	
	The forms of these matrices come from very few axioms, listed in \cite{Chams-200709}, and are not arbitrary. The zeros appear automatically and are necessary to exclude interactions not experimentally observed. Nonzero components are named after their coincidences with the fields and constants in the standard model. The first matrix is block diagonal and contains all the vector bosons. The second matrix contains Higgs terms. $D$ is a $192$ by $192$ matrix and acts on all $48$ known fermions. 
	
	The Fermionic part at \ref{eq-ac} justifies chosen names of fields and their coefficients as for nonzero components of $D_{AB}$. The first part of $D$ contains gauge fields as it does in the standard model; $B$, $V$, and $W$ stand for the $U(1)$, $SU(2)$, and $SU(3)$ gauge fields respectively \footnote{what we see here is an special case of a general theme, starting with a matrix algebra in the noncommutative geometry, the spectral action principle leads to a counterpart gauge theory.}. The second part is responsible for all the other fermion-fermion interactions which justifies the choice of names, Yukawa couplings $k^{i}$ and Higgs scalar fields $H^{a,b}$. In the trace part of action \ref{eq-ac} on the other hand, there is no fermionic field and the spectrum generates bosonic and scalar potentials which have the exact same form of standard model potential terms. In equation \ref{eq-poten} and what follows, we will study the scalar sector of the action. 
	
	To include antiparticles, we can double the algebra, and consequently the Hilbert space, by assuming the existence of a reality operator $J$ for the geometry as an axiom. This operator\footnote{It is evident that $J$ has the role here as the charge conjugate has in the standard model.} causes all the other operators to be the direct sum of two dependent parts which can be exchanged by the act of $J$. 
	
	The Dirac operator is however not simply the direct sum of fermionic and anti-fermionic parts. It is shown in \cite{Chams-2010} that only one off-diagonal element can be nonzero. This element therefore indicates a singlet that gives mass to a right handed fermion which is coinciding with a right handed neutrino in the standard model. Dirac operator of the whole space is therefore a 384 by 384 matrix as we noted before
	
\begin{align}\label{eq-dirmat}
D=\left(
\begin{matrix}
	D_{AB}&D_{AB^{\prime}}\\
	D_{A^{\prime}B}&D_{A^{\prime}B^{\prime}}
\end{matrix}
\right),
\qquad
&D_{A^{\prime}B}=\overline{D}_{AB^{\prime}}, \quad D_{A^{\prime}B^{\prime}}=\overline{D}_{AB}\\ \nonumber
&D_{AB^{\prime}}=\left(
\begin{matrix}
	\sigma&0..\\
	0..&0..
\end{matrix}\right)
\end{align}
	
	Having the above operator, both parts of the action (\ref{eq-ac}) are well defined. The fermionic part of action is containing fermion-gauge and fermion-Higgs interactions, plus terms coming from off-diagonal elements of $D$, which presents scalar-fermionic interactions absent in the standard model. Since $D_{AB^{\prime}}$ has only one nonzero element, only one of the fermions is involved with this new sigma-interaction and it is natural to call it right-handed neutrino \cite{Chams-2012}. 
\begin{equation}
	\langle \psi,D\psi \rangle=c \, \overline{\nu}_R \nu_R + C.C. + \text{fermionic and Yukawa interactions}
\end{equation} 
	
	Physically important geometrical information is also derivable from this operator and we need only to find coefficients introduced in (\ref{eq-seedew}) to identify the bosonic part of the action (\ref{eq-ac}). Calculations up to first three terms yield Einstein-Hilbert action along with Gauss-Bonnet terms, plus Higgs potential, $\sigma$ self-interaction, and $\sigma-H$ interaction. After proper redefinition of the fields, the scalar potential sector of (\ref{eq-ac}) is \cite{Chams-2010}:
\begin{equation}\label{eq-poten}
	V=\frac{1}{2} m^2_h H^2+\frac{1}{2} m^2_\sigma |\sigma|^2 + \frac{1}{4}\lambda_\sigma |\sigma|^4+ \frac{1}{4} \lambda_h H^4+ \frac{1}{2} \lambda_{h\sigma}|\sigma|^2 H^2.
\end{equation}
	
	We take $\sigma$ to be a complex singlet with two degrees of freedom. Although $H$ is a complex doublet with four degrees of freedom, the gauge symmetry allows us to gauge away three of them. The potential has local minimum which occurs when
\begin{align}
	\lambda_\sigma |\sigma|^2+\lambda_{h\sigma}H^2+{m^2_\sigma}=0, \quad \lambda_h H^2+\lambda_{h\sigma}|\sigma|^2+{m^2_H}=0
\end{align}
	and proposes the symmetry breaking, which we formulate with the following choices of the vacuum expectation values:
\begin{align}\label{eq-vev}
	H&=\left(\begin{matrix}0\\h+v\end{matrix}\right), \quad v={\langle h \rangle}_0\\ \nonumber
	\sigma&=w+\sigma_1+i \sigma_2, \quad w={\langle\sigma_1\rangle}_0.
\end{align}
	It is obvious from the above setting that the three scalars now mix and due to the $\sigma_1-\sigma_2$ symmetry, one massless Goldstone boson is expected to appear. After substituting (\ref{eq-vev}) into potential (\ref{eq-poten}), and diagonalizing the mass matrix of the square terms, the other two scalar masses turn out to be 
\begin{align}
	m^2_\pm=\left(v^2\lambda_h+\frac{w^2}{4}\lambda_\sigma\right)\left(1 \pm \left(1-\frac{v^2w^2\lambda_h \lambda_\sigma -v^2w^2(\lambda_{h\sigma})^2}{(v^2\lambda_h+\frac{w^2}{4}\lambda_\sigma)^2}\right)^{\frac{1}{2}}\right).
\end{align}
	It is believed that a highly massive right handed neutrino can be fitted in the standard model to explain neutrino oscillations. Such a neutral particle is only able to gain mass from a singlet scalar field. In the model described above this mechanism is appearing naturally. The price of this treatment is of course to have a new scalar which is supposed to be highly massive.  Here we have another massless field added to the picture which appears since $\sigma$ is a complex field. We therefore suppose $w$ to be much greater than $v$ and we get
\begin{align}\label{eq-higgs}
	M= w \sqrt{\frac{\lambda_\sigma}{2} }, \qquad m_h =v \sqrt{2\lambda_h}\sqrt{1 - \frac{\lambda_{h\sigma}^2}{ \lambda_h \lambda_\sigma}}.
\end{align}
	The smaller one is responsible for the Higgs mass and is modified by the factor of $\sqrt{1 - \frac{\lambda_{h\sigma}^2}{\lambda_h \lambda_\sigma}}$ due to the presence of the scalar field. It is remarkable that noncommutative geometry not only predicts the singlet field and its potential terms, but also relates, in the unification scale, the scalar couplings to other parameters such as Yukawa couplings and the unified gauge coupling \cite{Chams-2010}. Having those relations, we will start from unification and vary all the free parameters to probe the implications of this formula for the Higgs mass.
	
\subsection{Running of the renormalization group equations}\label{sec-running}
	Having the model described in section \ref{sec-model}, one can find the effective potential and renormalization group equations in some loop order and run them to explore high energy scales. There are however two free parameters here. The neutrino Yukawa coupling and the Higgs self-coupling. Knowing the Higgs mass now, the value of Higgs self-coupling is determined in the pure standard model as
	\[\lambda_{h}(M_{z})=\frac{(125.5)^2}{2(246.2)^2}=0.1299.\]
	In models with extra scalars though, there is a see-saw  mechanism which determines the Higgs mass and the value of this coupling is not determined even when the mass is measured.
	
\begin{figure}[h]
		\includegraphics[width=0.49\linewidth]{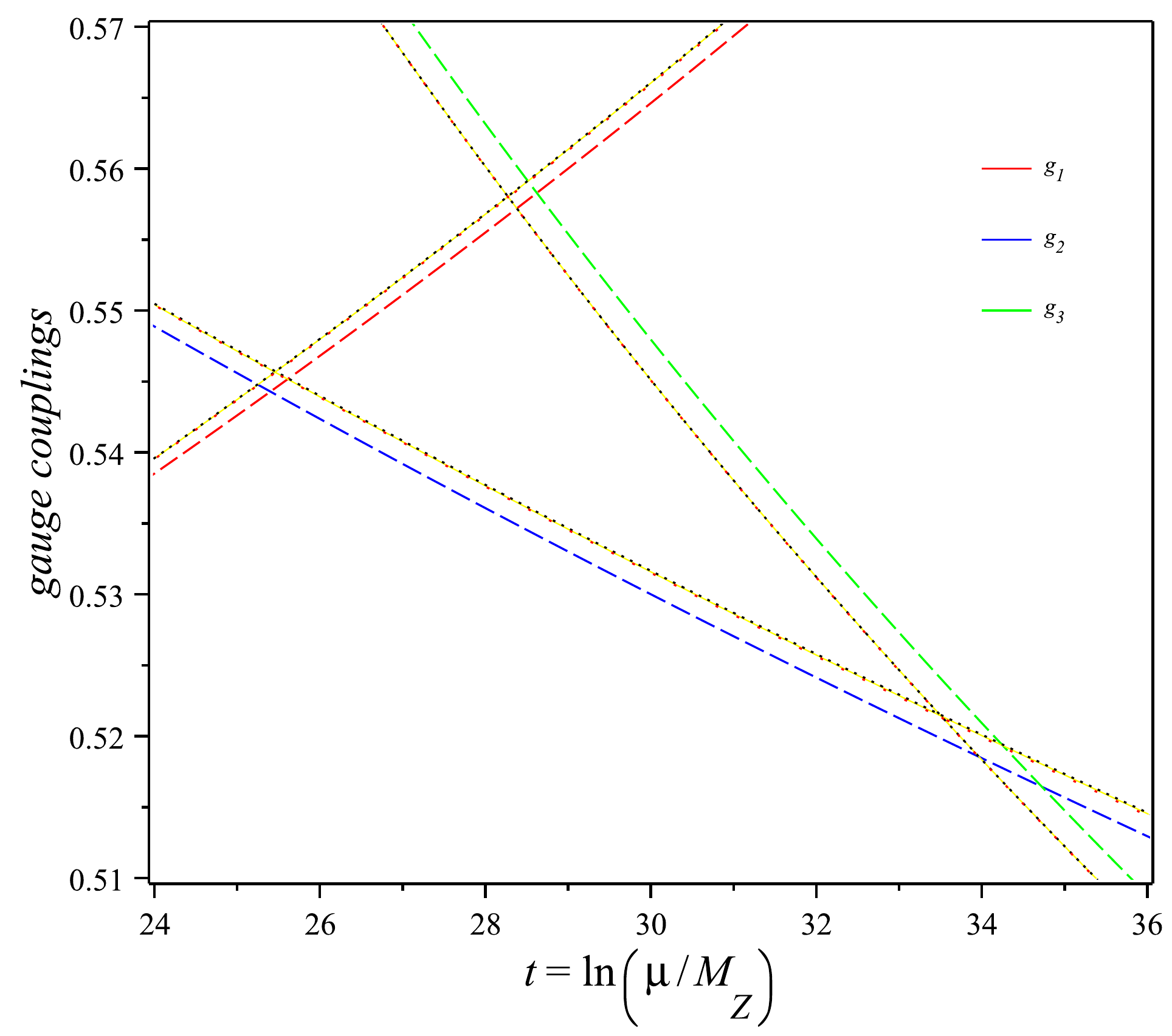}
		\includegraphics[width=0.49\linewidth]{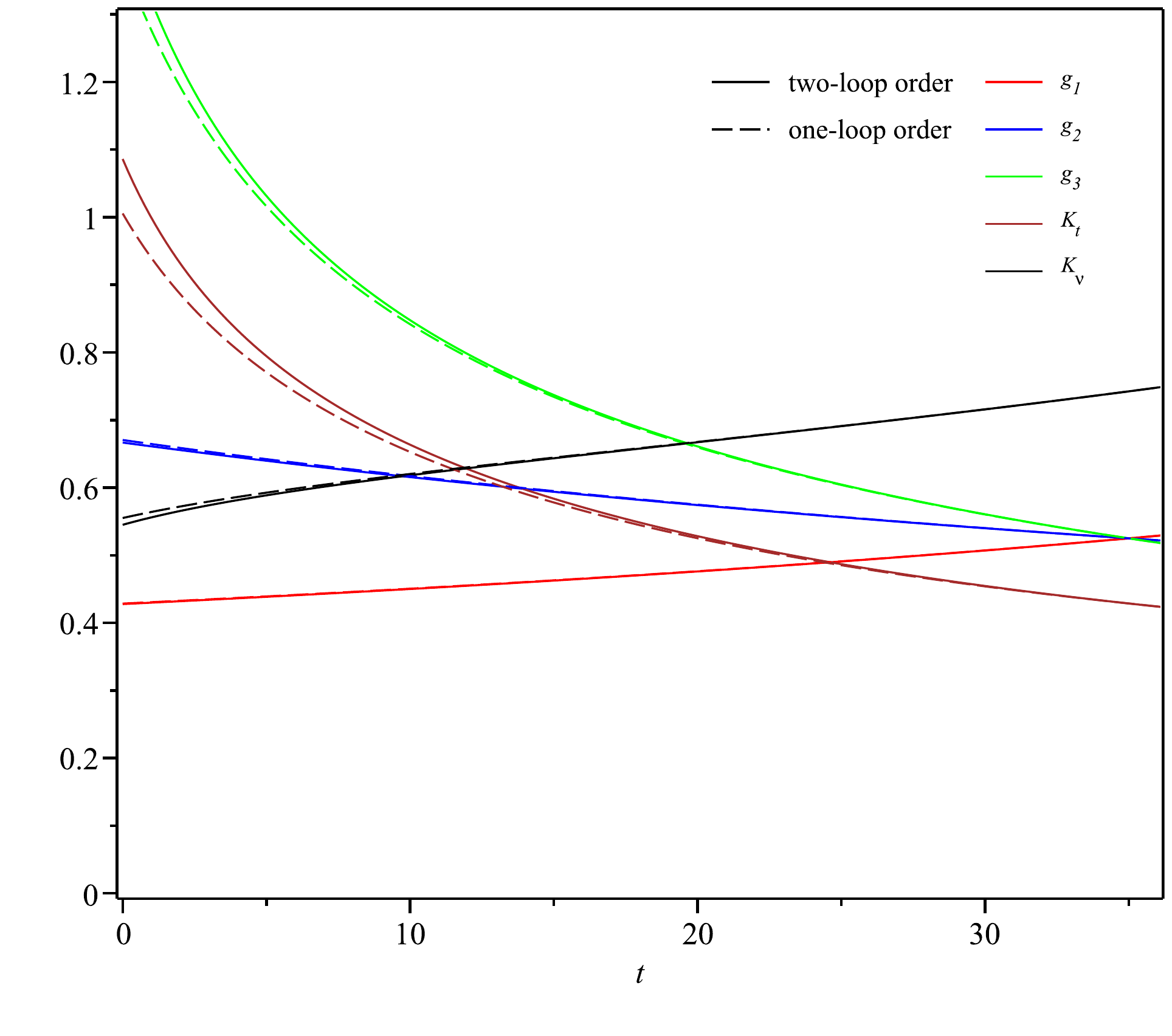}
		\caption{The behavior of gauge couplings at the unification scale. Dashed lines are indicating the evolution of standard model couplings up to one-loop corrections. Yellow solid lines show the situation is slightly better when two-loop corrections are also taken into account. The black dotted lines are for RG equations up to three loops for the standard model. The black and yellow lines are so close that their separation cannot be distinguished in this diagram. This difference is from the same order of errors that  experimental uncertainties create when we run the equations upward. In all cases, the corrections coming from a real or complex scalar field added to the standard model is negligible. Red dotted lines have two-loop corrections of the complex scalar field, in the model described in section \ref{sec-model}, and include three-loop corrections of all the other parameters. Yet again it matches with two-loop corrections suggesting that higher orders are not going to make the situation any better. The graph on the left compares one-loop RGEs with two-loop equations for gauge and Yukawa couplings when we start at the same points at high energies and follow them toward experimental values. Again adding a singlet doesn't creates meaningful changes.}			
		\label{fig-unification}
\end{figure}

	We used SARAH which is a Mathematica package to derive two-loop RG equations (\cite{sarah-2008}) for this model and presented the results in appendix \ref{ap1}. It is clear from RG equations that the extra field cannot correct the gauge couplings evolutions and therefore is not going to help the couplings to meet in exactly one point (Figure \ref{fig-unification}). That is because the scalar field potential terms are quadratic and their couplings appear only in two-loop corrections of Yukawa couplings evolutions, which themselves enter just in two-loop corrections of the gauge couplings. The latter is due to the Yukawa interaction of the particles with square of the singlet. Figure \ref{fig-unification} also shows that the added singlet field, no matter is it complex or real, does not cause meaningful changes in evolution of Yukawa couplings. However, two-loop corrections shift them for about ten percent if we follow their evolutions to very different energy scales.
	
	Though replacing the real scalar field with a complex field has no remarkable implications on the gauge couplings evolutions, it can cause noticeable consequences for the field couplings as there are new Feynman diagrams between them when we add the imaginary component. Choosing acceptable initial values and running RG equations, including two-loop effects, show that this difference is meaningful. As it is clear from Figure \ref{fig-runup}, starting from the same points, the couplings behave differently at very high energies. These couplings indicate Higgs mass through the relation (\ref{eq-higgs}) which can be sensitive to small variations of the couplings.	
	
\section{Top quark and Higgs masses at low energies} \label{sec-3}
\begin{figure}
		\begin{subfigure}{\linewidth}
			\centering
			\includegraphics[width=0.9\linewidth]{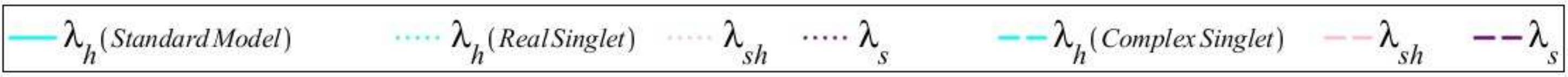}
		\end{subfigure}
		\begin{subfigure}[t]{0.49\linewidth}
			\includegraphics[width=1\linewidth]{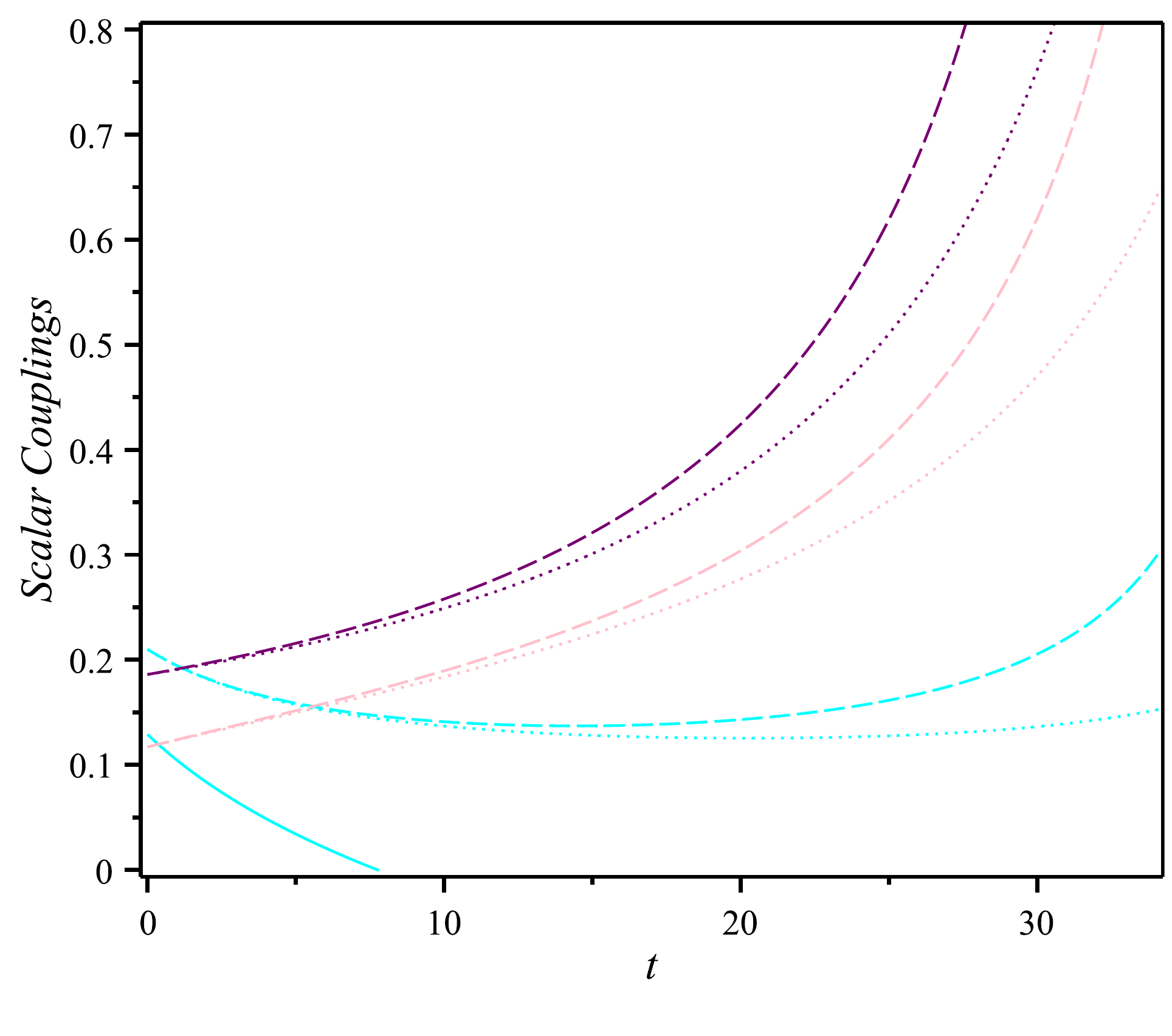}
			\caption{}
			\label{fig-runup}
		\end{subfigure}
		\begin{subfigure}[t]{0.49\linewidth}
			\includegraphics[width=1\linewidth]{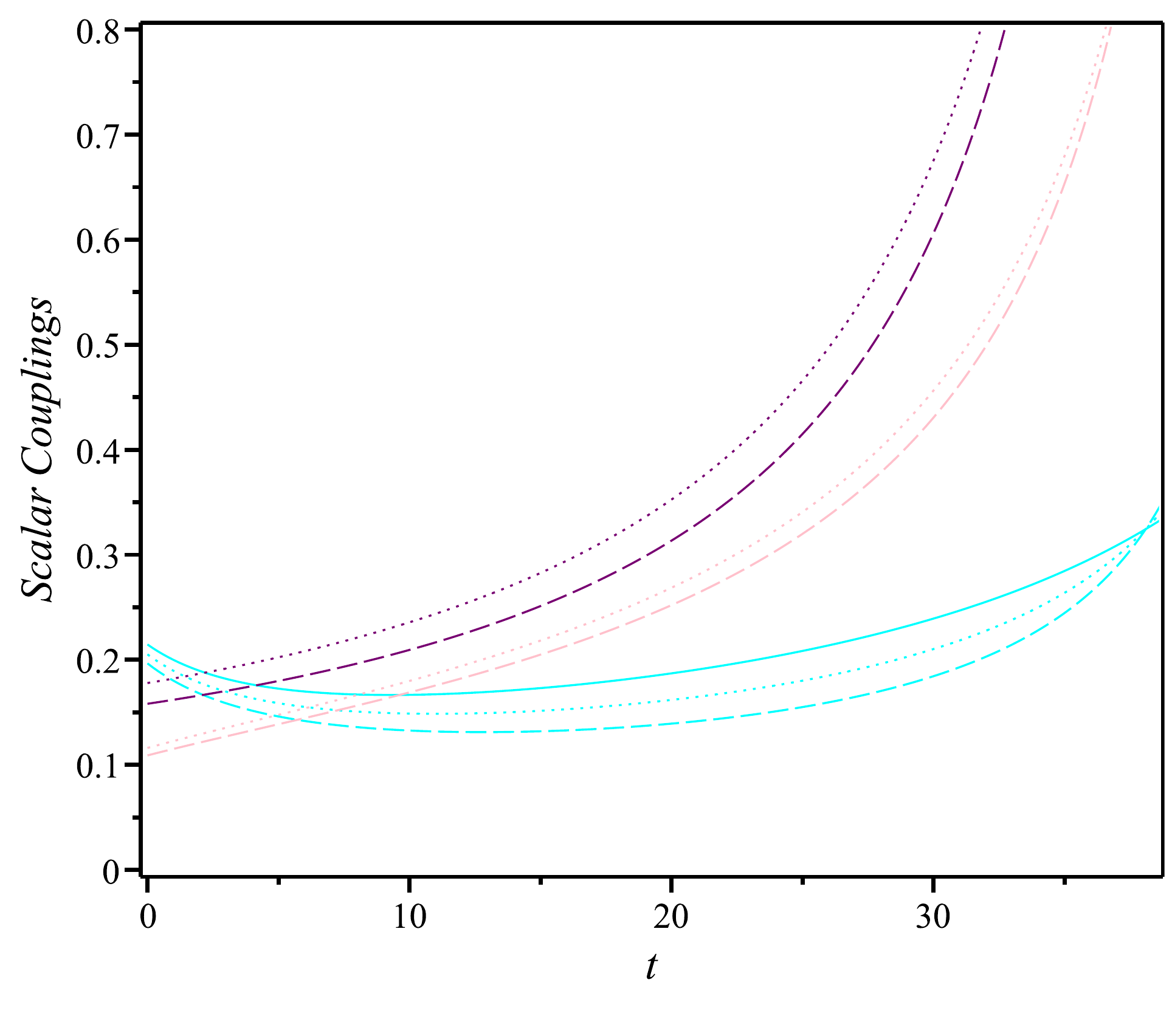}
			\caption{}
			\label{fig-rundown}
		\end{subfigure}			
		\caption{Running of the couplings incorporating two-loop corrections, toward unification (a) or from unification (b). Solid lines are for the SM, dots are for when the real scalar singlet is present, and dashed lines are for the case that the singlet field is complex. In case of SM, except for the neutrino Yukawa coupling, the initial values are coming from experiments. In all the cases, the initial values for experimentally unknown couplings are discussed in the next chapter; when we run from unification and look for the best fits.}
\end{figure}

	In our approach, we run RG equations downward from the unification scale. The advantage is that the spectral model predicts initial conditions at high energies, and relates all the Yukawa couplings to the unified gauge coupling $g$. Interestingly, the scalar couplings are also not free parameters at the unification energy, instead they are determined by both $g$ and the ratio of neutrino and  top quark Yukawa couplings \cite{Chams-2007}. We choose the approach of \cite{Chams-2012} and, for simplicity, define the ratio $n=(\frac{k^\nu}{k^t})^{\frac{1}{2}}$ at the unification scale. $n$ is one of the free parameters of the model which can be fixed, then running this along with other parameters causes predictions for the physical quantities at the experimental arena. As discussed before however, the unification scale itself and the value of gauge couplings at this scale is not predicted. Figure \ref{fig-rundown} shows the evolution of all the parameters in different scenarios. There is about ten percent difference in the values of the couplings at low energies between real and complex models. Since the effects of higher orders of loop corrections are negligible, the  difference we see here does worth investigating. Another encouraging fact is that in \cite{Chams-2012} the effects of the scalar field couplings were shown to be able to save the model after the Higgs small mass discovery.
		
	The other observation which justifies our consideration reveals itself when we compare two-loop and one-loop equations. Whether complex or real singlet is added to the Lagrangian, the scalar couplings get modified for about ten percent at low energies and as noted before this can in principle dramatically modify results of equation (\ref{eq-higgs}).
	
	It is notable that if we use initial values coming from the spectral action, it is not possible to run the minimal standard model from unification scale and find the experimentally acceptable mass of the Higgs particle at low energies. Trivially the reason is that these initial conditions imply unification of the gauge couplings which does not happen for minimal standard model. In \cite{Chams-2012} however, the authors showed initial conditions and RG equations are consistent with the low Higgs mass for when the added scalar field coming from spectral model is real. Nevertheless, spectral action imposes no restrictions on the singlet field. In this section, we consider both complex and real scalars and use two-loop RG equations to incorporate higher order corrections and asses the importance of loop corrections. We also study prediction of the theory for top quark mass.
	
	Our main result in this section is that the assumption of existence of a new scalar field and the predictions of the spectral model at unification are consistent with the known masses of top quark and Higgs. For simplicity we neglect lighter particles. As it was noted in section \ref{sec-model} however, the neutrino is assumed to play a significant role since it has a Yukawa coupling and its mass comes from a see-saw mechanism. The method is straight forward; We assume the initial conditions predicted in \cite{Chams-2007}. Then we run the equations supposing $n$, $g$, and $U$ are free parameters. It hands us couplings values at low energies. Then it is possible to find the best values for these three parameters by minimizing the errors between the result masses and experimentally known values at low energies. The fact that these errors exist and are more than experimental uncertainties is very important and we will discuss it in the next section.  
	
	The other important aspect of the situation is to compare two-loop and one-loop corrections, as we are comparing real scalar and complex scalar fields. Up to one-loop, there is no remarkable change in top quark mass if we replace the real scalar with a complex one. However, the two-loop corrections differentiate top quark mass in these two cases. This differentiation is still one order of magnitude smaller than the current observational uncertainties. The situation is different for the Higgs mass as it depends directly to the scalar couplings (eq. \ref{eq-higgs}). 
	
\begin{figure}
		\includegraphics[width=0.48\linewidth]{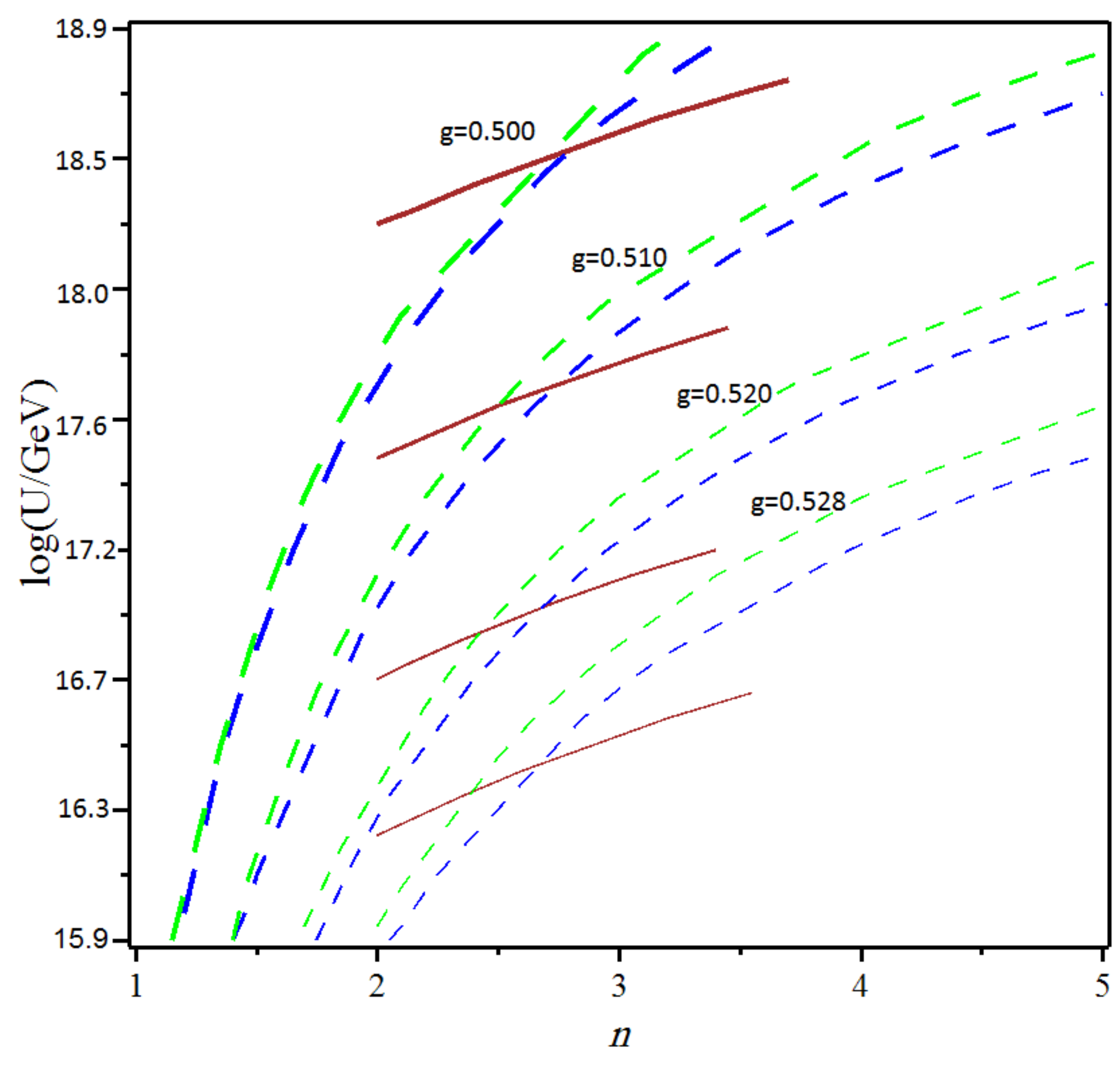}		
		\includegraphics[width=0.48\linewidth]{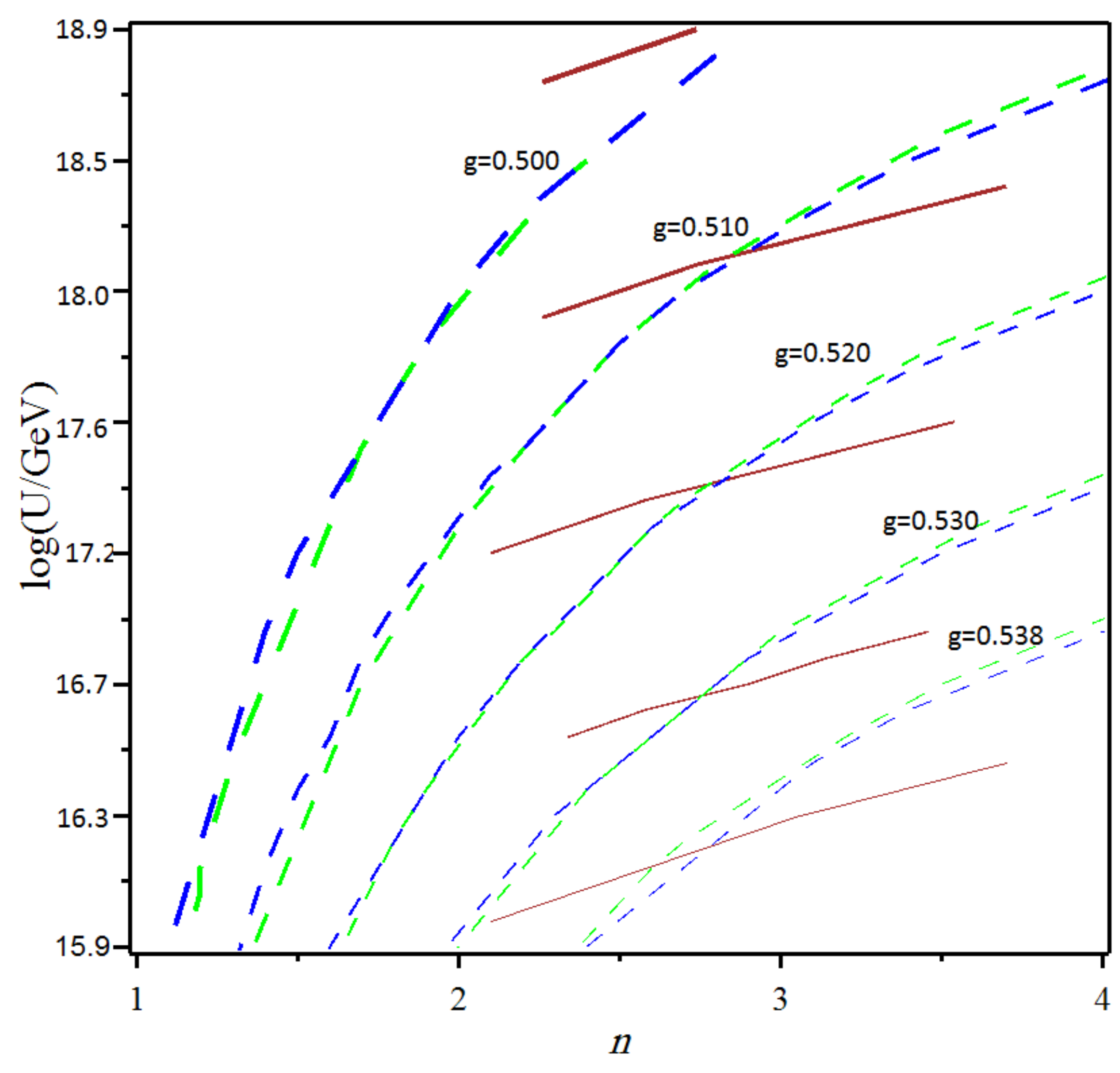}
		\caption{Each line shows suitable choices of unification scale and $n$ value at this scale, in order to revive experimental values of particle masses at low energies. Each set of three lines are for a specific $g$ value and are illustrated with a particular thickness. The left hand side diagram incorporates two-loop corrections while the diagram on the right has only one-loop corrections. It can be inferred from diagrams that within a reasonable range of $g$, the lines associated with top quark, solid brown lines, and Higgs, dashed lines, always have a collision point. Therefore suitable $n$ and $U$ can be always found to assure the low energy values for the Higgs and top quark masses. This is true for both real and complex cases which are distinguished by blue and green lines respectively.}
		\label{fig-untwoloop}
\end{figure}

	Our considerations shows that there is a rather short range for $n$ and $g$ that everything fits together. This happens for a $U$, unification scale, varying between $2\times10^{16} GeV$ and $5\times10^{18} GeV$. In figure \ref{fig-untwoloop} the lines indicate what initial values are acceptable to meet the correct particle masses at law energies. It turns out that for a reasonable g, the correct choices for n and U always exist to fit the Higgs and top quark masses simultaneously in low energies within the experimentally acceptable values. 
	
\begin{figure}
		\includegraphics[width=0.48\linewidth]{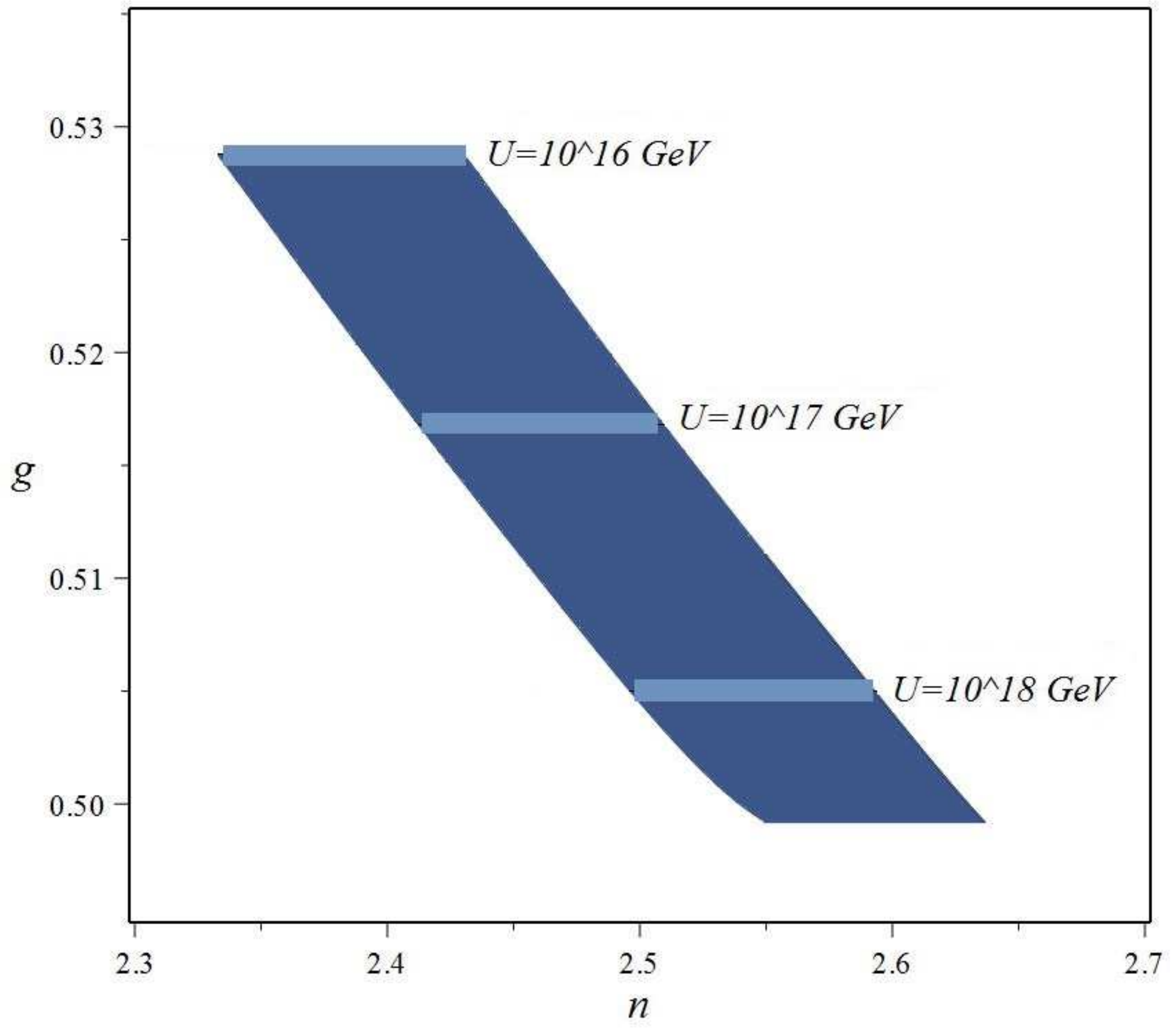}
		\includegraphics[width=0.48\linewidth]{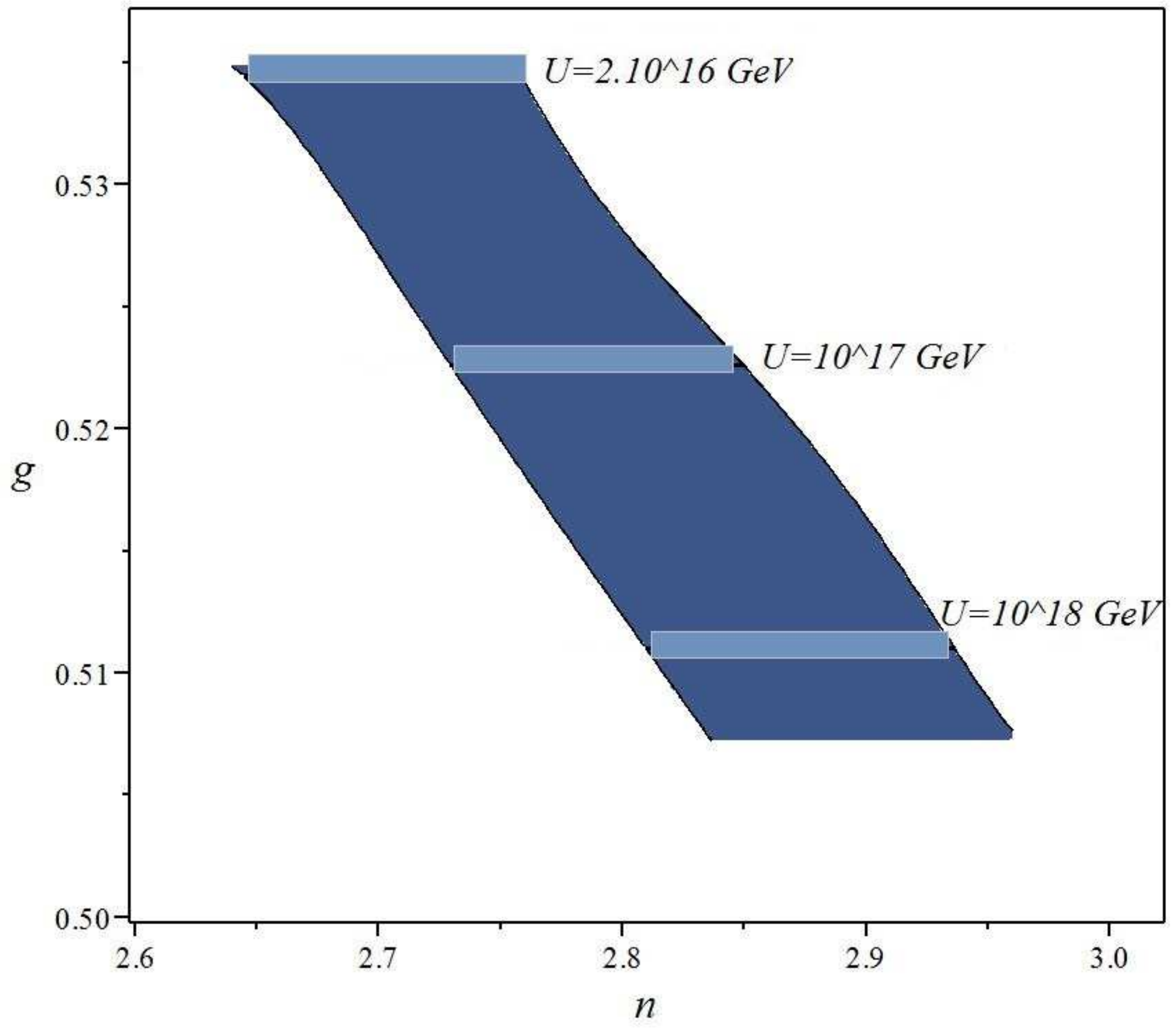}	
		\caption{At any unification scale, there is a small window of choices for unified gauge coupling, $g$, and the root of neutrino and top quark Yukawa couplings ratio, $n$, which lead to consistent low energy particle masses with experimental values. Two-loop corrections, left diagram, make the choices a little more restricted.}
		\label{fig-g-n}	
\end{figure}

	To illustrate even more, we show possible choices for $g$ and $n$ at unification energy in figure \ref{fig-g-n}. The colored strips in two diagrams show all the choices which lead to retrieving particle masses at low energies, incorporating one-loop or two-loop corrections. As we noted before however, the correct choice of unification scale is depending on $g$ and $n$. To give some examples, the small window of correct choices of $g$ and $n$ for three different unification energies are indicated by lighter colors on the strips.
	
	Up to two loop corrections, the suitable $n$ is obtained to be around $2.7$ for the real scalar and around $2.5$ for the complex scalar case which means that at the unification scale, Yukawa coupling of neutrino is around $6$ times bigger than the top quark coupling.

\subsection{comparing the complex and the real cases}
	We saw that for both scenarios (complex or real singlets), it is possible to find acceptable initial conditions. On the other hand, again in both cases, gauge values deviations at low energies do not fit within the experimental uncertainties. Yet, the situation is slightly better in the complex case for $g_{3}$ and $g_{2}$. For any $g$ at unification, $U$ is about $0.2$, and $n$ is about $0.28$ higher in the real case compared with the complex one.
	
	For the standard model alone, best quantities are: $g\sim0.49$  and $u\sim39$ When the scalar (complex or real) is added, up to one-loop, $g\sim0.52$ and $u\sim35$ and up to two-loop, $g\sim0.53$ and $u\sim32$ end to the best results.
	
\section{Implications on the vacuum instability}
\begin{figure}[h]
		\centering
		\includegraphics[width=0.5\linewidth]{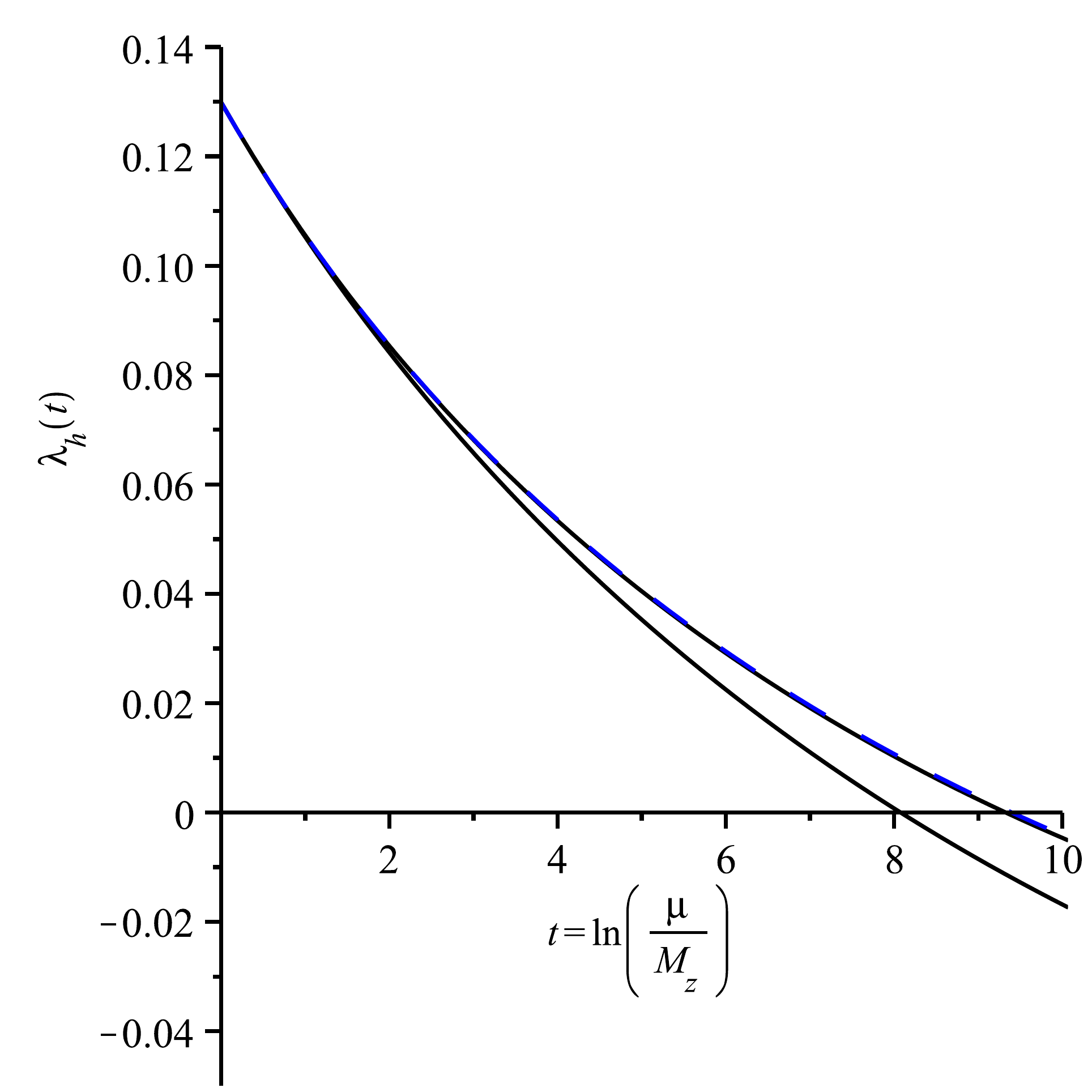} 
		\caption{After the Higgs discovery, all of the initial values are known for the standard model parameters and one can follow the evolution of Higgs self-coupling. The lower line has only one-loop corrections. The line shifts to the right when two-loop effects are added to RG equations, and the tunneling time increases consequently. The blue dashed line includes three-loop corrections and suggests that going to higher orders will not improve the situation.}
		\label{fig-higgscoSM}
\end{figure}

	In the standard model, the observed masses of Higgs and top quark imply effective potential of the Higgs field to become unstable at high energies. This can be seen, in the tree level, by the fact that the Higgs self-coupling changes its sign at some energy scale below the unification. For the standard model itself, one can use the renormalization group equations up to some order and find the point at which $\lambda_{h}$ changes its sign. It turns out that this happens at the energy scale of order $10^{6} GeV$ \footnote{To find this result, in addition of all the known couplings of SM, we take into account the Yukawa of neutrino which is around $0.5$. It shifts the instability to lower energies, however later when we add the singlet and all the parameters of the model, the instability goes to much higher energies.} which is much smaller than the unification scale. Figure \ref{fig-higgscoSM} shows two-loop corrections have an effective role to make the situation better while three-loop corrections are too small to have any significance. Thus, we do not expect higher order corrections to resolve this issue. 
	
\begin{figure}[ht]	
	\begin{subfigure}[t]{0.49\linewidth}
		\includegraphics[width=\linewidth]{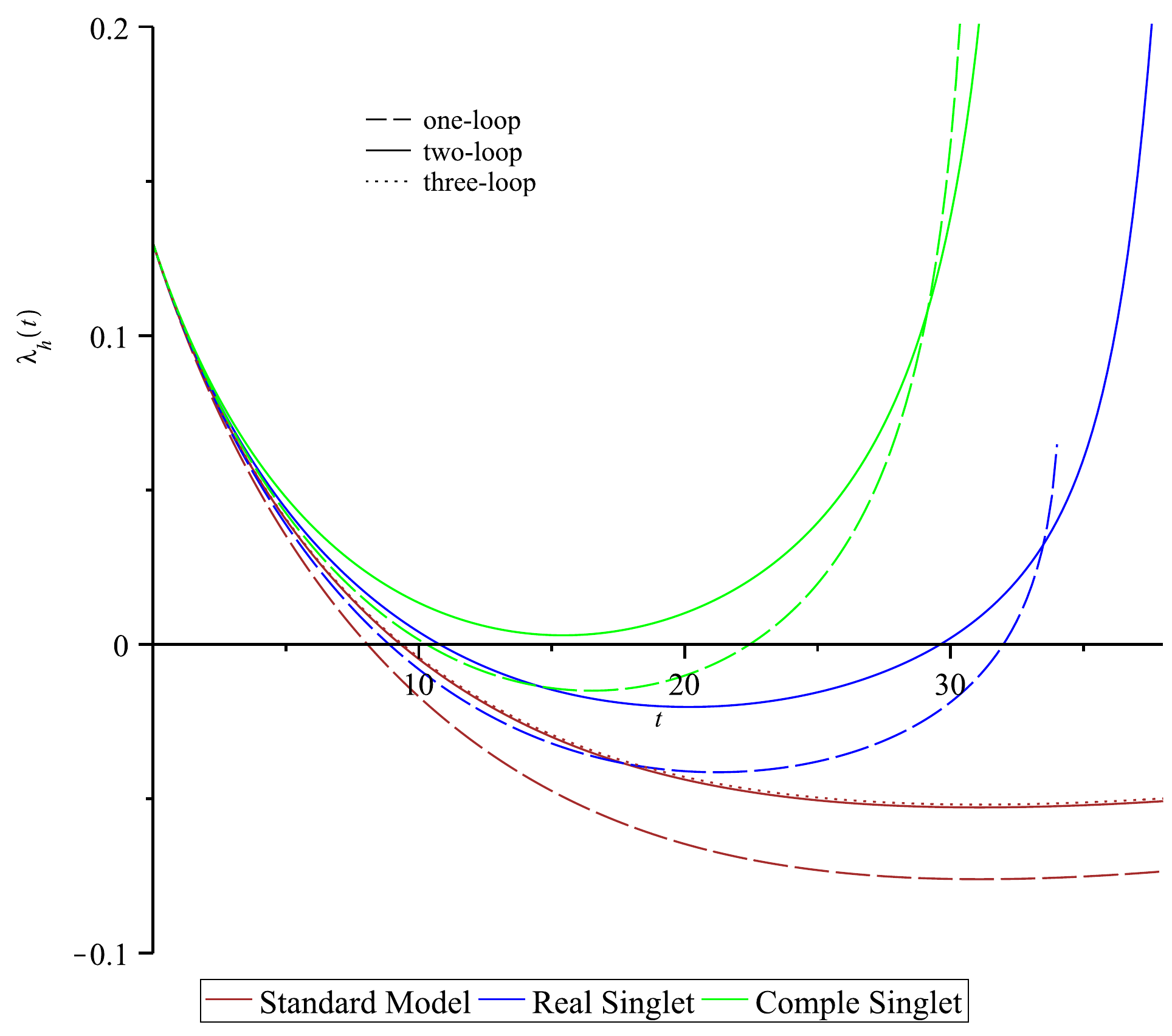}
		\caption{The initial values used to draw this diagram are not realistic. However it shows that addition of new fields can in principle have effects more than higher loop corrections. The errors of these lines at high energies due to the experimental uncertainties of the initial values are less than ten percent of loop effects.}
		\label{fig-higgsinstability}
	\end{subfigure}
	\hspace{5pt}
	\begin{subfigure}[t]{0.49\linewidth}
		\includegraphics[width=\linewidth]{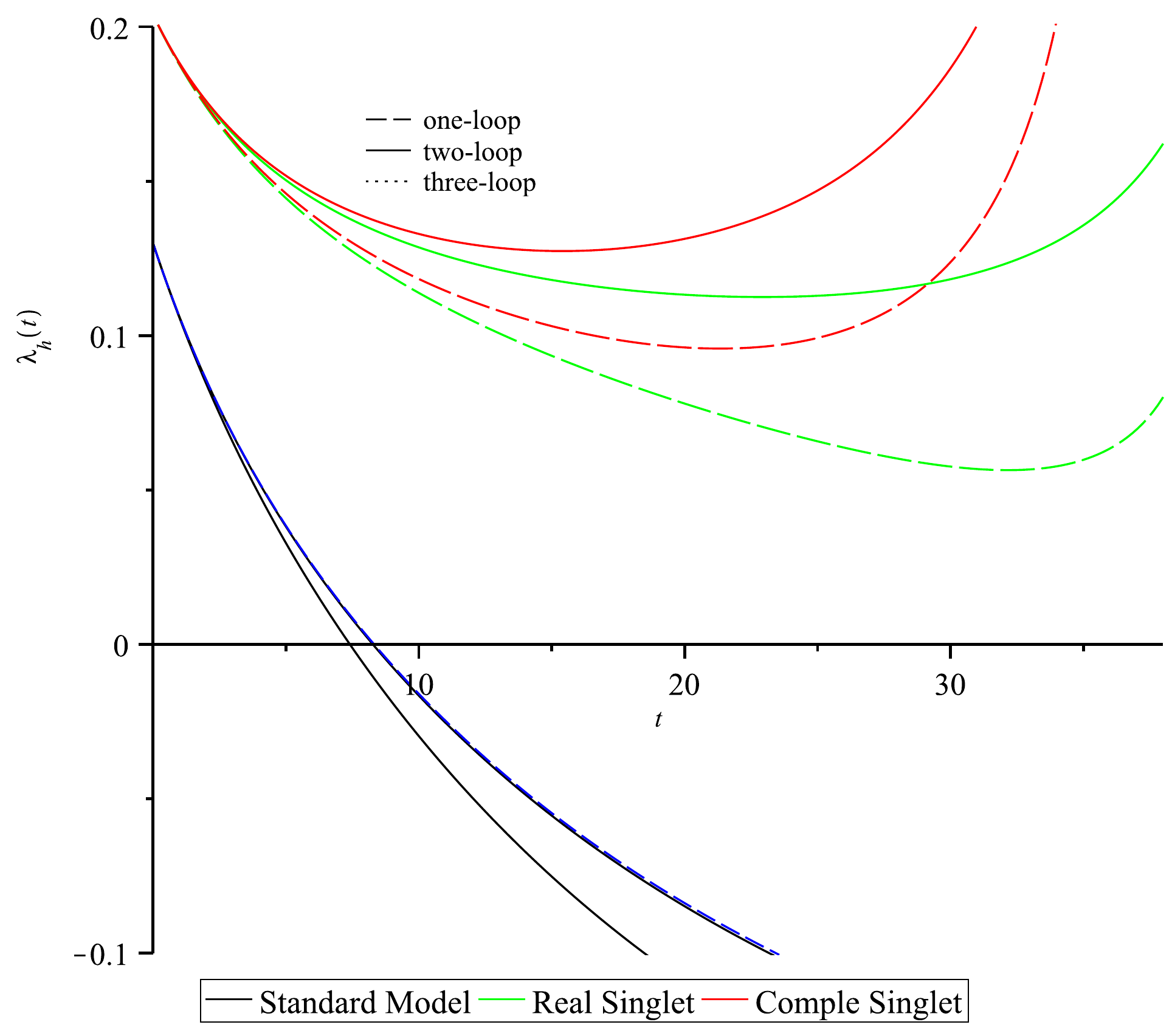}
		\caption{Comparison between the behavior of Higgs self-coupling in different scenarios. In scalar extended standard model, the coupling is not determined with Higgs mass and could have a greater initial value which might save the potential from being instable.}
		\label{fig-runhiggsup}
	\end{subfigure}
	\caption{}
\end{figure}

	With an additional scalar field, it is interesting to see what happens for the effective potential. There are two new couplings associated with the scalar quadratic term and its interaction with the Higgs in the model described earlier. These two couplings along with the Higgs self-coupling are only constrained by the masses of the Higgs and the supposedly heavily massive singlet. Therefore there are not enough known initial conditions and one cannot run the renormalization group equations from low energies. It is however useful to investigate whether this additional field could in principle modify the equations as much as needed in order to cure instability. A straightforward investigation shows that the addition of a complex field could in principle cure the equations (Figure \ref{fig-higgsinstability}). As noted before, it is especially important due to the fact that higher loop corrections are not being expected to save the potential. 
	
	As we saw in the previous sections, there are a number of predictions at high energy scales in spectral approach which suggest to start from the unification and run RG equations downward. Doing this gives ideas about the acceptable range of values for couplings; particularly this is useful for the extra couplings which we have no clue about their magnitudes as they are not constrained with the current experimental data. One result is that the Higgs self-coupling is stronger than in the pure standard model, and this pushes the instability of the effective potential to higher energies. It does not affect the Higgs mass because of the see-saw mechanism between Higgs and new scalar. Figure \ref{fig-runhiggsup} illustrates what happens when we use such initial conditions. All the couplings in the potential are now positive all along the way up to unification scale and the potential is expected to be stable. In this respect both real and complex scalar models behave desirably. 

\section{Conclusion}
	The noncommutative model introduced in section \ref{sec-model} adds some familiar extra features to the standard model, for example a new singlet field with quadratic potential, higher powers of geometrical invariants, and prediction of gauge couplings unification at high energies. Our considerations in this paper show that starting from the unification point predicted by the theory, it is possible to revive both top quark and Higgs masses. We however witnessed that there is a deviation for the gauge coupling values at experimental arena which is of the same order of deviation of the gauge couplings in the standard model at unification scale. Comparing these errors we conclude that the complex singlet field makes the theory slightly better than the pure standard model or when a real scalar is added; however, the full treatment is not possible. 
	
	Comparing the results of two-loop corrections and near to leading order, for three-loop, shows that there is no hope for loop corrections to contribute in a significant way. We believe the root of all of such inconsistencies goes back to the issue of gauge couplings not meeting at one point and therefore lack of a true unification. Equivalently in noncommutative geometry approach there is unification, but the price in the simple version that we considered here was that one could not fully revive the gauge couplings at low energies. Yet, the little change toward better results with this minimal change in the settings of the standard model might urge us to investigate other more generalized models derived from noncommutative geometry principles. 
	
	The spectral action approach coming from the noncommutative geometry point of view, however, does not uniquely lead to the model we considered here. Further investigations showed in 2014 that imposing generalized versions of Heisenberg uncertainty relations leads to Pati-Salam model as the most general possible outcome of this approach \cite{Chams-2013}. The model we considered here is the simplest special case of that general theory. The Pati-Salam model has a rich content of beyond SM fields that might help the situation and will be the subject of our further investigations.
	
\section*{Appendix}
	\appendix
\section{2-loop RGEs for complex singlet extended standard model}\label{ap1}
	
	Here we present 2-loop renormalization group equations of the complex singlet extended standard model with right-handed neutrino. These equations are derived using SARAH package for Mathematica \cite{sarah-2008}. The equations are consistent with the literature \cite{Costa:2014qga,Chen:2012faa,Gonderinger:2012rd}.
	
	\begin{dgroup*}[style={\footnotesize},spread={5pt}]
		
		\begin{dmath*}
			\frac{d g_{1}}{dt} ={\frac {41\,{g_{1}}^{3}}{160\,{\pi }^{2}}}+\frac {{g_{1}}^
				{3}}{12800\,{\pi }^{4}} \Big( -15\,{K_{\nu }}^{2}-85\,{K_{t}}^{2}+199\,{g_{1}}^{2}+135\,{
				g_{2}}^{2}+440\,{g_{3}}^{2} \Big) ,
		\end{dmath*}
		
		\begin{dmath*}
			\frac{d g_{2}}{dt}=-{\frac {
					19\,{g_{2}}^{3}}{96\,{\pi }^{2}}}+\frac {{g_{2}}^{3}}{7680\,{\pi }^{4}} \Big( -15\,{K_{
					\nu }}^{2}-45\,{K_{t}}^{2}+27\,{g_{1}}^{2}+175\,{g_{2}}^{2}+360\,{g_{3
			}}^{2} \Big) ,
		\end{dmath*}
		
		\begin{dmath*}
			\frac{d g_{3}}{dt}=-{\frac {7\,{g_{3}}^{3}}{16\,{
						\pi }^{2}}}+\frac {{g_{3}}^{3}}{2560\,{\pi }^{4}} \Big( -20\,{K_{t}}^{2}+11\,{g_{1}}^{2
			}+45\,{g_{2}}^{2}-260\,{g_{3}}^{2} \Big) ,
		\end{dmath*}
		
		\begin{dmath*}
			\frac{d K_{\nu}}{dt}=\frac {K_{\nu }}{16 {\pi }^{2}} \Big( -9/20 \, g_{1}^{2}-9/4\,{g_{2}}^{2}+3\,{K_{t}}^{2}+5/2\,{K_{\nu }}^{2}
			\Big) +\frac {1}{256\,{\pi }^{4}} \left( 1/40\, \Big( 21\,{g_{1}
			}^{4}-54\,{g_{1}}^{2}{g_{2}}^{2}-230\,{g_{2}}^{4}+240\,{\lambda _{h}}^
			{2}+80\,{\lambda _{{\it sh}}}^{2}+5\, \left( 17\,{g_{1}}^{2}+45\,{g_{2
			}}^{2}+160\,{g_{3}}^{2} \right) {K_{t}}^{2}+15\, \left( {g_{1}}^{2}+5
			\,{g_{2}}^{2} \right) {K_{\nu }}^{2}-270\,{K_{t}}^{4}-90\,{K_{\nu }}^{
				4} \Big) K_{\nu }+{\frac {{K_{
							\nu }}^{3}}{80}} \Big( -60\,{K_{\nu }}^{2}-540\,{K_{t}}^{
				2}+279\,{g_{1}}^{2}+675\,{g_{2}}^{2}-960\,\lambda _{h} \Big)  \right) ,
		\end{dmath*}
		
		\begin{dmath*}
			\frac{d K_{t}}{dt}=\frac {K_{t}}{16 {\pi }^{2}} \left( -
			{\frac {17\,{g_{1}}^{2}}{20}}-9/4\,{g_{2}}^{2}-8\,{g_{3}}^{2}+9/2\,{K_
				{t}}^{2}+{K_{\nu }}^{2} \right) +\frac {1}{256\,{\pi }^{4}} \left( 
			\frac {K_{t}}{
				600} \left( 1187\,{g_{1}}^{4}-270\,{g_{1}}^{2}{g_{2}}^{2}-3450\,{g_
				{2}}^{4}+760\,{g_{1}}^{2}{g_{3}}^{2}+5400\,{g_{2}}^{2}{g_{3}}^{2}-
			64800\,{g_{3}}^{4}+3600\,{\lambda _{h}}^{2}+1200\,{\lambda _{{\it sh}}
			}^{2}+75\, \left( 17\,{g_{1}}^{2}+45\,{g_{2}}^{2}+160\,{g_{3}}^{2}
			\right) {K_{t}}^{2}+225\, \left( {g_{1}}^{2}+5\,{g_{2}}^{2} \right) {
				K_{\nu }}^{2}-4050\,{K_{t}}^{4}-1350\,{K_{\nu }}^{4} \right) + \left( {\frac {223\,{g_{1}}^{2}}{80}}+{\frac {135\,{g_{2}}^{2}
				}{16}}+16\,{g_{3}}^{2}-12\,\lambda _{h}-{\frac {27\,{K_{t}}^{2}}{4}}-9
			/4\,{K_{\nu }}^{2} \right) {K_{t}}^{3}+3/2\,{K_{t}}^{5} \right) ,
		\end{dmath*}
		
		\begin{dmath*}
			\frac{d \lambda _{h}}{dt}=\frac {1}{16 {\pi }^{2}} \left( {\frac {27\,{g_{1}}^{4}}{200}}+{\frac 
				{9\,{g_{1}}^{2}{g_{2}}^{2}}{20}}+{\frac {9\,{g_{2}}^{4}}{8}}-9/5\,{g_{
					1}}^{2}\lambda _{h}-9\,{g_{2}}^{2}\lambda _{h}+24\,{\lambda _{h}}^{2}+
			4\,{\lambda _{{\it sh}}}^{2}+12\,\lambda _{h}\,{K_{t}}^{2}+4\,\lambda 
			_{h}\,{K_{\nu }}^{2}-6\,{K_{t}}^{4}-2\,{K_{\nu }}^{4} \right) +
			\frac {1}{256\,{\pi }^{4}} \left( -{\frac {3411\,{g_{1}}^{6}}{2000}}-{
				\frac {1677\,{g_{1}}^{4}{g_{2}}^{2}}{400}}-{\frac {289\,{g_{1}}^{2}{g_
						{2}}^{4}}{80}}+{\frac {305\,{g_{2}}^{6}}{16}}+{\frac {1887\,{g_{1}}^{4
					}\lambda _{h}}{200}}+{\frac {117\,{g_{1}}^{2}{g_{2}}^{2}\lambda _{h}}{
					20}}-{\frac {73\,{g_{2}}^{4}\lambda _{h}}{8}}+{\frac {108\,{g_{1}}^{2}
					{\lambda _{h}}^{2}}{5}}+108\,{g_{2}}^{2}{\lambda _{h}}^{2}-312\,{
				\lambda _{h}}^{3}-40\,\lambda _{h}\,{\lambda _{{\it sh}}}^{2}-32\,{
				\lambda _{{\it sh}}}^{3}+ \left( -{\frac {171\,{g_{1}}^{4}}{100}}-9/4
			\,{g_{2}}^{4}+{\frac {45\,{g_{2}}^{2}\lambda _{h}}{2}}+80\,{g_{3}}^{2}
			\lambda _{h}-144\,{\lambda _{h}}^{2}+1/10\,{g_{1}}^{2} \left( 63\,{g_{
					2}}^{2}+85\,\lambda _{h} \right)  \right) {K_{t}}^{2}-{\frac {{K_{\nu }}^{2}}{200}} \Big( 
			18\,{g_{1}}^{4}+15\,{g_{1}}^{2} \left( 4\,{g_{2}}^{2}-20\,\lambda _{h}
			\right) +150\,{g_{2}}^{4}-300\,{g_{2}}^{2}\lambda _{h}+9600\,{
				\lambda _{h}}^{2} \Big) -8/5\,{g_{1}}^{2}{K_{t}
			}^{4}-32\,{g_{3}}^{2}{K_{t}}^{4}-3\,\lambda _{h}\,{K_{t}}^{4}-\lambda 
			_{h}\,{K_{\nu }}^{4}+30\,{K_{t}}^{6}+10\,{K_{\nu }}^{6} \right),
		\end{dmath*}
		
		\begin{dmath*}
			\frac{d \lambda _{sh}}{dt} =
			\frac {\lambda _{{\it sh}}}{160\,{\pi }^{2}} \Big( 60\,{K_{t}}^{2}+20\,{K_{\nu }}^{2}
			-9\,{g_{1}}^{2}-45\,{g_{2}}^{2}+120\,\lambda _{h}+80\,\lambda _{{\it 
					sh}}+80\,\lambda _{s}  \Big) -
			\frac {1}{102400\,{
					\pi }^{4}}\lambda _{{\it sh}}\, \left( -1671\,{g_{1}}^{4}-450\,{g_{1}}^{2
			}{g_{2}}^{2}+3625\,{g_{2}}^{4}-5760\,{g_{1}}^{2}\lambda _{h}-28800\,{g
				_{2}}^{2}\lambda _{h}+24000\,{\lambda _{h}}^{2}-480\,{g_{1}}^{2}
			\lambda _{{\it sh}}-2400\,{g_{2}}^{2}\lambda _{{\it sh}}+57600\,
			\lambda _{h}\,\lambda _{{\it sh}}+17600\,{\lambda _{{\it sh}}}^{2}+
			38400\,\lambda _{{\it sh}}\,\lambda _{s} \left( t \right) +16000\,
			\left( \lambda _{s} \left( t \right)  \right) ^{2}-100\, \left( 17\,{
				g_{1}}^{2}+45\,{g_{2}}^{2}+160\,{g_{3}}^{2}-288\,\lambda _{h}-96\,
			\lambda _{{\it sh}} \right) {K_{t}}^{2}-100\, \left( 3\,{g_{1}}^{2}+15
			\,{g_{2}}^{2}-96\,\lambda _{h}-32\,\lambda _{{\it sh}} \right) {K_{
					\nu }}^{2}+5400\,{K_{t}}^{4}+1800\,{K_{\nu }}^{4} \right) ,	
		\end{dmath*}
		
		\begin{dmath*}
			\frac{d \lambda _{s}}{dt} =\frac 
			{1}{16\, {\pi }^{2}} \left( 8\,{\lambda _{{\it sh}}}^{2}+20\, \left( \lambda _{s} \left( t
			\right)  \right) ^{2} \right)+\frac {1}{256\,{\pi }^{4}}
			\left( {\frac {48\,{g_{1}}^{2}{\lambda _{{\it sh}}}^{2}}{5}}+48\,{g_{
					2}}^{2}{\lambda _{{\it sh}}}^{2}-64\,{\lambda _{{\it sh}}}^{3}-80\,{
				\lambda _{{\it sh}}}^{2}\lambda _{s} \left( t \right) -60\, \left( 
			\lambda _{s} \left( t \right)  \right) ^{3}-48\,{\lambda _{{\it sh}}}^
			{2}{K_{t}}^{2}-16\,{\lambda _{{\it sh}}}^{2}{K_{\nu }}^{2} \right).
		\end{dmath*}

	\end{dgroup*}
	
\section*{Acknowledgment}	
	The author is grateful to Professor Ali Chamseddine for suggesting the problem and also all the insightful discussions and suggestions. He also would like to thank Jad El Hajj and Hasan Hammoud for their help on working with the Linux packages.
	
	\bibliography{ref-complexsinglet}{}
	\bibliographystyle{plain}
	
\end{document}